\documentclass[9pt,onecolumn,twoside]{osajnl}
\journal{ol} % Choose journal (ao,jocn,josaa,josab,ol,optica,pr)

\setboolean{shortarticle}{true}
\usepackage{color,soul}
\usepackage{lineno}
\usepackage{color}

\title{3D inhomogeneous self-accelerating
beams}

\author[1,2]{Jing Pan}
\author[1,2]{Hao Wang}
\author[3,*]{Yijie Shen}
\author[1,2,$\star$]{Xing Fu}
\author[1,2,$\dagger$] {Qiang Liu}

\affil[1]{Key Laboratory of Photonic Control Technology (Tsinghua University), Ministry of Education, Beijing 100084, China}
\affil[2]{State Key Laboratory of Precision Measurement Technology and Instruments, Department of Precision Instrument, Tsinghua University, Beijing 100084, China}
\affil[3]{Optoelectronics Research Centre, University of Southampton, Southampton SO17 1BJ, UK}

\affil[*]{Corresponding author:
Y.Shen@soton.ac.uk}
\affil[$\star$]{Corresponding author: fuxing@mail.tsinghua.edu.cn}
\affil[$\dagger$]{Corresponding author: qiangliu@mail.tsinghua.edu.cn}

\begin{abstract}
We propose and generate a new class of structured light fulfilling quantum-like coherent states based on a set of circular Airy vortex modes. Such coherent-state wave packets possess strong focus with both radial and angular self-accelerations, which exploit more general 3D inhomogeneous velocity control with global spatial symmetry of multilayer rotation akin to galactic kinematics, as termed galaxy waves. Galaxy waves are endowed with new degrees of freedom to control strong focusing and acceleration of 3D structured light, promising numerous applications in optical trapping, manufacturing, and nonlinear optics.
\end{abstract}
\begin{document}
\maketitle
Airy beams have received continuous attention in recent years~\cite{efremidis2019airy}. Due to unique self-accelerating characteristic and curved parabolic geometry~\cite{siviloglou2007observation}, Airy beams were widely used in filamentation~\cite{polynkin2009curved,polynkin2009filamentation}, particle manipulation~\cite{baumgartl2008optically}, and imaging~\cite{jia2014isotropic}. In particular, circular Airy beams evolve along the curved track radially, and thus are also termed auto-focusing Airy beams~\cite{panagiotopoulos2013sharply}. Vortex beams are also concerned closely for their rotating wavefronts and orbital angular momenta (OAM) which benefit a myriad of applications such as particle trapping, communications, imaging and quantum
entanglement~\cite{shen2019optical,forbes2021structured}. Introducing vortex phase into the circular Airy beams, circular Airy vortex beams (CAVBs) have both rotating and self-accelerating characteristics for more flexible applications~\cite{davis2012abruptly}. Specially, energy of the CAVBs converges to the center upon propagation before the focal region, and the evolutions of their inner and outer phase distributions are inhomogeneous~\cite{2012Propagation,chen2015propagation}. Nonetheless, the CAVB only shows concentric-circle intensity distribution and lacks general geometry control to extend applications, especially in particle manipulation. To deal with this, the tornado wave is proposed as 
superposition of two CAVBs to break the uniformity of angular light intensity distribution, which brings rotating angular velocity features to self-accelerating structured light~\cite{brimis2020tornado}. However, the rotating angular velocity of this wave packet changes only upon propagation in one dimension. Is it possible to create and control accelerating structured light in 3D? In fact, in order to control 3D geometric modes, quantum-analogue coherent states with general SU(2) symmetry have been widely used to design complex spatial wave packets with stable patterns and controllable trajectory~\cite{shen2021rays,shen2020,shen2021creation}. Many exotic structured beams as SU(2) coherent states but using diverse bases have been designed for customized geometric patterns, based on Hermit-Laguerre-Gaussian modes~\cite{shen20202}, Bessel modes~\cite{liang2020generation}, and Ince-Gaussian modes~\cite{wang2021unify}. However, extant modes designed by this method cannot achieve self-acceleration and lack inhomogeneous angular velocity evolution in 3D space, which limits the extension of applications.

In this Letter, we design a new class of self-accelerating structured light whose wave packets have 3D inhomogeneous angular velocity evolution, in terms of the propagation dimension and transverse two dimensions characterizing multilayer rotations of intensity peaks in diverse angular velocities, akin to galactic kinematics, thus termed galaxy waves. The galaxy wave is crafted by CAVB superposition fulfilling the form of SU(2) coherent state. SU(2) in mathematics provides rich tunable parameters for mode control, which enable us to tune both angular velocities of layered rotations and global symmetry of a galaxy wave. The experimental control of galaxy waves largely increases the flexibility of light shaping and accelerating in multiple spatial dimensions so that offers new insight in applications such as optical trapping, manufacturing, and nonlinear optics.

\begin{figure}
	\centering
	\includegraphics[width=0.45\linewidth]{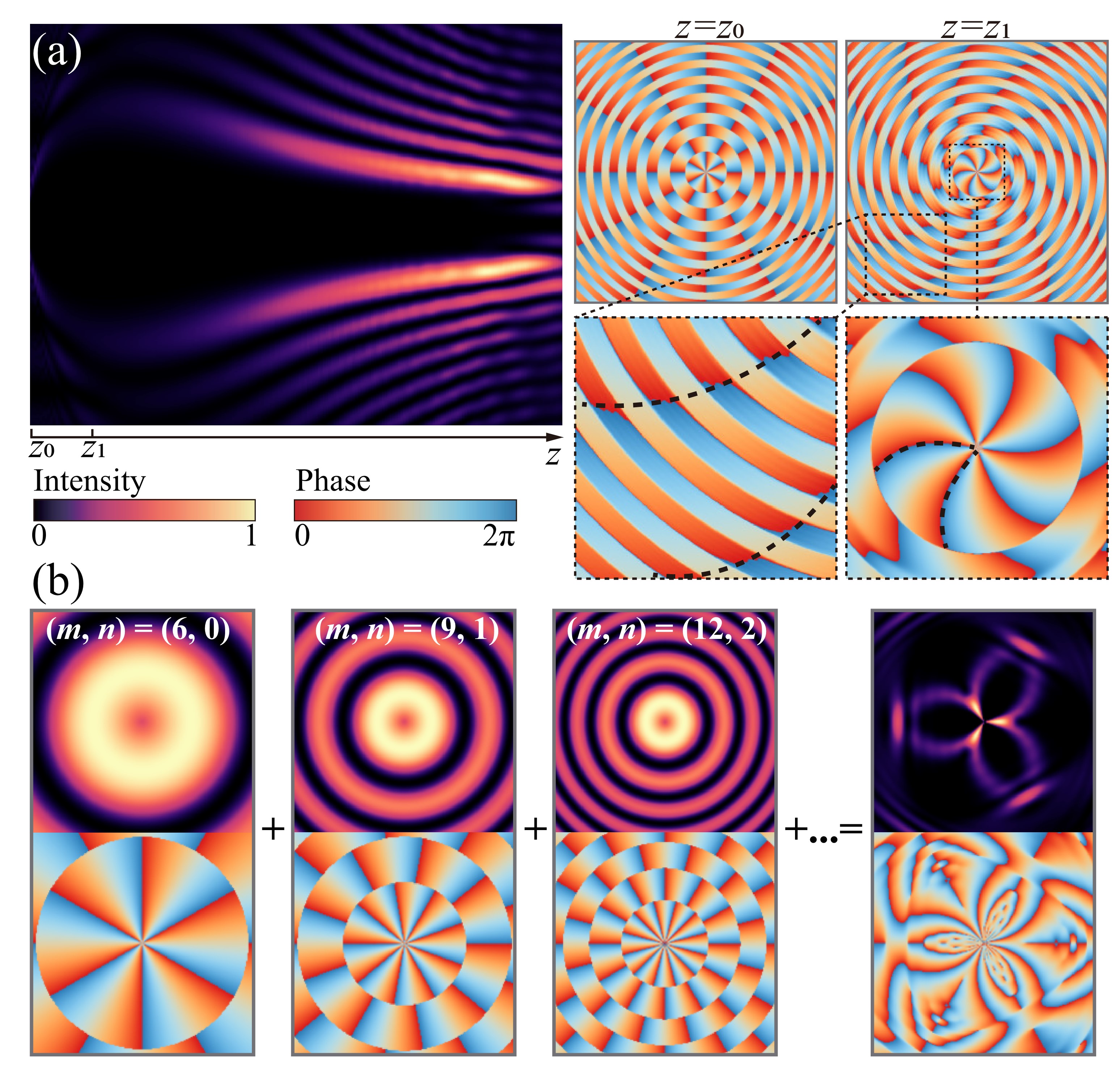}
	\caption{(a) $x$-$z$ longitudinal section of intensity distribution of the CAVB with $m=6$ and $n=0$ and corresponding phase distributions at $z_{0}$ and $z_{1}$. (b) Transverse intensity and phase distributions of various CAVBs with different $m$ and $n$ at $z_{0}$ with generated galaxy wave shown as the right.}
	\label{fig:1}
\end{figure}

First, it's necessary to introduce CAVBs briefly. CAVBs have both the self-accelerating characteristic of circular Airy beams and rotating phase characteristic of vortex beams, which can be expressed as~\cite{brimis2020tornado}. 
\begin{equation}
		\Phi_{m,n}(r,\theta)={\rm Ai}(\frac{r_{0}-r}{R/\alpha_{m,n}}){\rm exp}(a\cdot\frac{r_{0}-r}{R/\alpha_{m,n}}){\rm exp}(im\theta)
		\label{eq:1}
\end{equation}
where ${\rm Ai}(\cdot)$ represents Airy function, $(r,\theta)$ are the polar coordinates in transverse plane, $a$ is the exponential truncation factor, $r_{0}$ is the radius of the primary Airy ring, $m$ is the OAM topological charge, and $R$ is the scaling factor. Here, a new parameter $\alpha_{m,n}$ is introduced for greater flexibility in tuning the radial light field distribution, which represents the $n^{\rm th}$ zero point of the $m^{\rm th}$-order Bessel function. When this new parameter is set to a fixed number which is irrelevant to $m$ and $n$, the expression degenerates to the traditional one. CAVBs have intriguing properties, one of which is autofocusing. Specifically, energy flows to the center with evolution so that the intensity of center regions increases sharply within a certain propagation distance. This can be seen from Fig.~\ref{fig:1} (a) which clearly shows the $x$-$z$ longitudinal section of intensity evolution distribution of the CAVB with $(m,n)=(6,0)$. Moreover, CAVBs also have peculiar phase distribution, especially inhomogeneous phase evolution of the inner and outer light field regions, as shown in the right part of Fig.~\ref{fig:1}(a). The above two insets show phase distributions of the CAVB with $(m,n)=(6,0)$ at $z_{0}$ and $z_{1}$, corresponding to the marked positions on the left intensity figure. The differences between the phase evolutions of the inner and outer parts are clearly shown in the below zoom-in panels. 

With the introduced parameter $\alpha_{m,n}$, a CAVB can be expanded to a set of CAVBs under the control of $m$ and $n$, which provides larger potential for mode superposition. Here we propose a new structured light family, termed as the galaxy waves, which fulfill SU(2) coherent state mathematical superposition rule and leverage CAVB modes as base elements. A galaxy wave is expressed as:
\begin{equation}
	\begin{split}
		\Psi_{m_0,n_0}^{(p,q,N,\phi)}(r,\theta)=\sum_{k=0}^{N}{\binom{N}{k}}^{1/2}{\rm exp}(ik\phi)\Phi_{m_{0}+qk,n_{0}+pk}(r,\theta)
		\label{eq:2}
			\end{split}
		\end{equation}
where $\phi$ is the phase factor in the SU(2) coherent state, and $\Phi_{m_{0}+qk,n_{0}+pk}$ represents a CAVB mode with $m=m_{0}+qk$ and $n=n_{0}+pk$ in Eq.~\ref{eq:1}. $m_{0}$ and $n_{0}$ are initial mode orders, $p$ and $q$ are interval numbers, and $N$ is related to the number of superposed modes. Here, we take $m_{0}=6$, $n_{0}=0$, $p=1$, $q=3$, $N=5$ as an example, which includes modes with indices $(m,n)=(6,0),(9,1),(12,2),(15,3),(18,4),(21,5)$. By superposing these CAVBs shown in Fig.~\ref{fig:1} (b), generated galaxy wave is shown on the right, in which inner and outer radial bi-layer peak regions follow angular 3-fold symmetry. The symmetry is decided by the $q$ index. 

\begin{figure}
	\centering
	\includegraphics[width=0.5\linewidth]{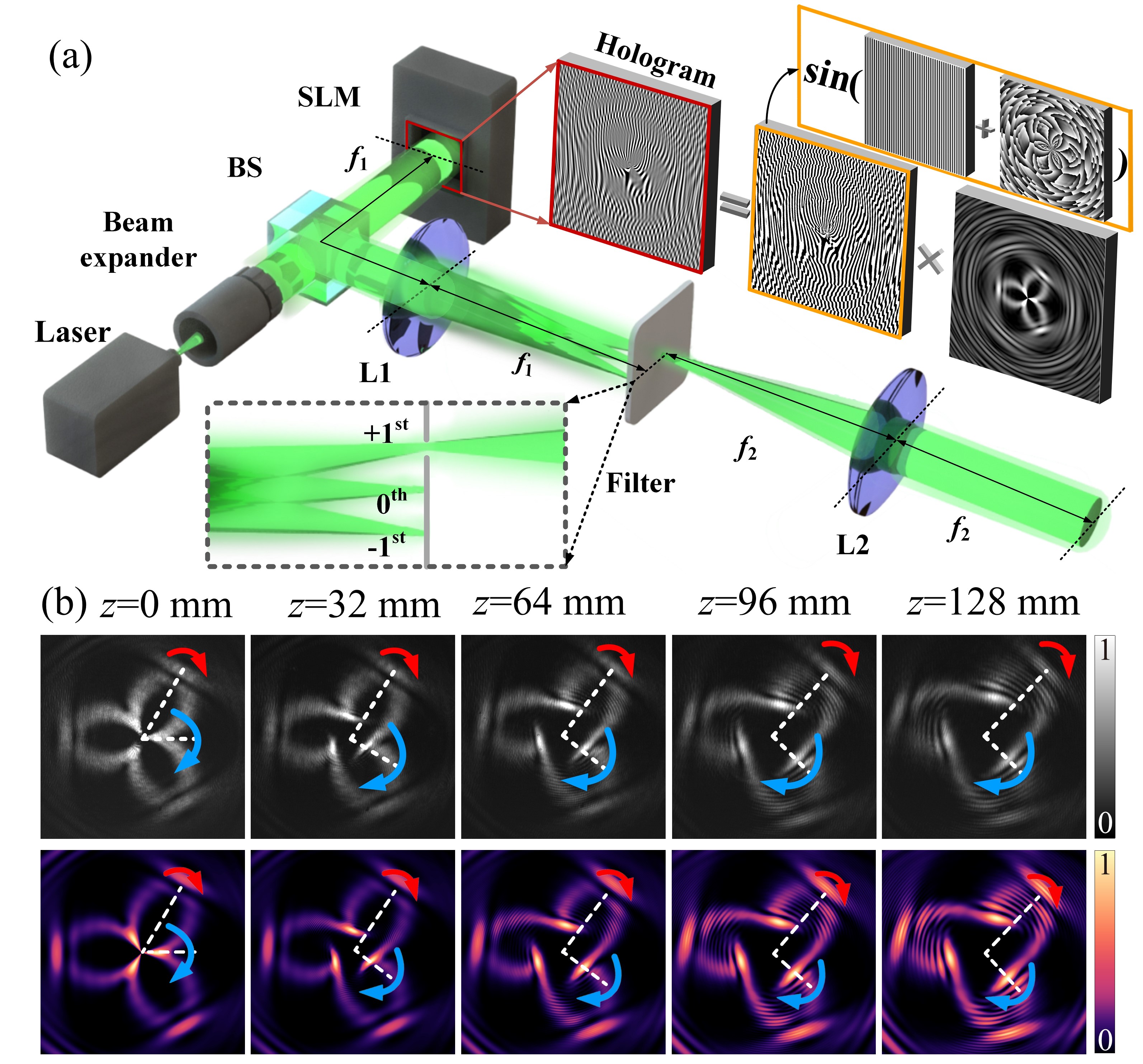}
	\caption{(a) Experimental setup. BS: beam splitter; L1: a lens with focal length $f_{1}$: 50~mm; L2: a lens with focal length $f_{2}$: 75~mm. (b) The experimental (upper) and simulated (lower) results of transverse pattern evolved from $z=0$ to $z=128$~mm.}
	\label{fig:2}
\end{figure}

The galaxy wave is experimentally generated based on digital hologram method using a spatial light modulator (SLM). The experimental setup is shown in Fig.~\ref{fig:2} (a). A 532~nm beam from the laser is expanded to 12.8~mm diameter on the SLM, and illuminates the loaded phase holograms to carry information of the designed modes. See the enlarged mask in the red box next to the SLM in Fig.~\ref{fig:2} (a). The hologram is generated by multiplying the amplitude pattern of the on-demand mode and the phase component together. The phase component is the superposition of phase distribution of the on-demand mode and the blazed grating. After modulation by the holograms and filtering, the $+1^{st}$-order diffracted beam becomes the galaxy wave and is recorded by a charged-coupled device (CCD) at different propagation distance $z$. (The focal plane of L2 is seen as the plane with $z=0$~mm.) 

\begin{figure}
	\centering
	\includegraphics[width=0.6\linewidth]{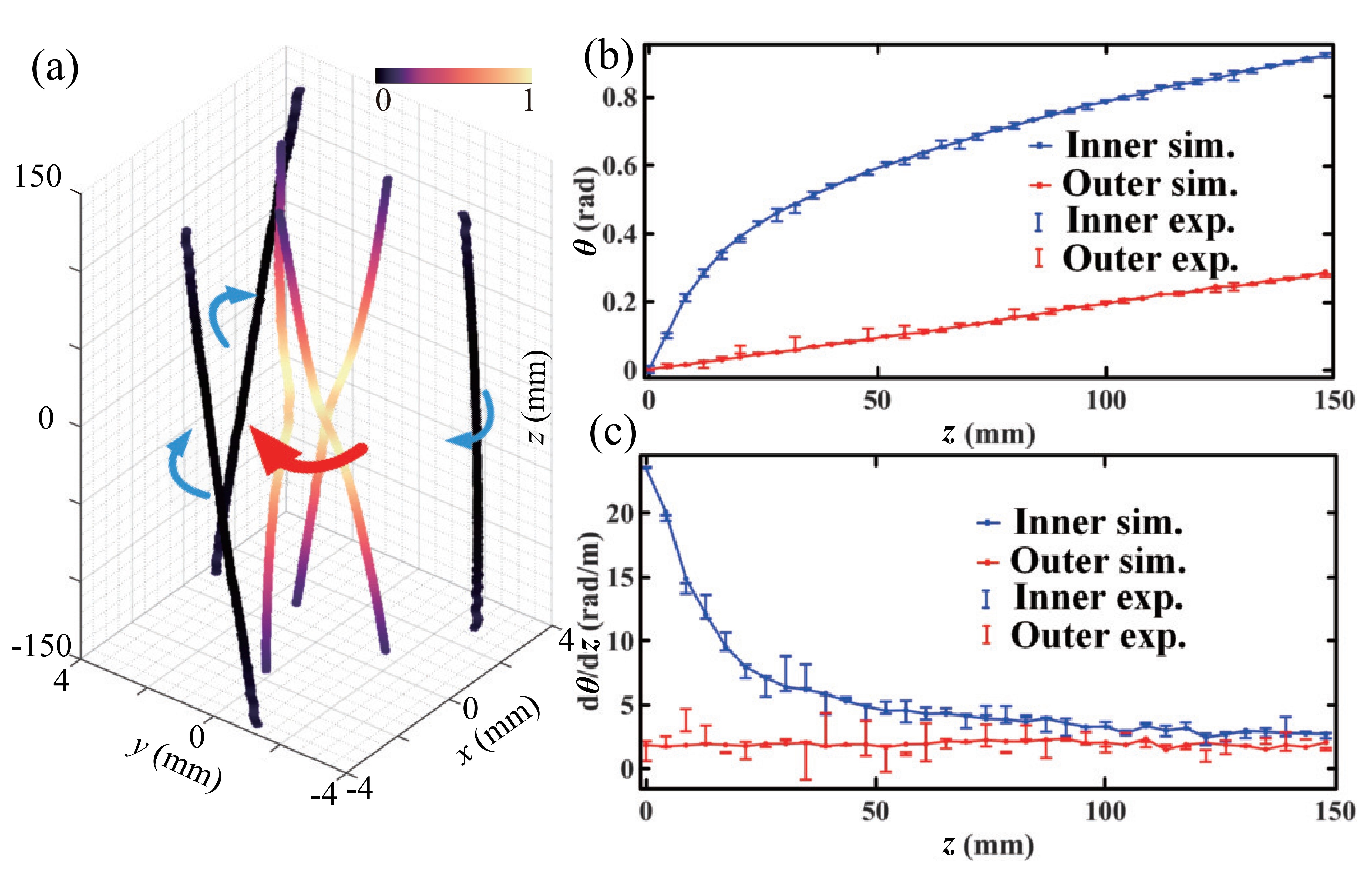}
	\caption{(a) Theoretical intensity peak trajectories of the galaxy wave. (b,c) Experimental and simulated results of angular position (b) and velocity (c) of the intensity peaks upon propagation. The inner and outer peak regions correspond to blue and red simulated lines, respectively; the error bars evaluate the errors to find peak average angular positions (b) and average angular velocities (c) in experiment.}
	\label{fig:3}
\end{figure}

The experimental intensity distributions from $z=0$~mm to $z=128$~mm are shown in Fig.~\ref{fig:2} (b), which match well with the simulated results below. It can be seen that peak regions of inner and outer light fields rotate in different angular velocities upon propagation, which can be credited to the inhomogeneous phase distribution evolution of CAVBs. Intensity peak regions are extracted as shown in Fig.\ref{fig:3} (a) to clearly showcase the self-accelerating evolution and rotation features upon propagation. In Fig.\ref{fig:3} (a), energy of light field flows to the center twice when propagation distance $z$ increases, resulting in intensity bi-peak distribution upon propagation. The calculated Poynting vector is further shown in {\color{blue}{Visualization~1}} with the range $z\ge0$, and the intensity evolution of the other side of $z=0$ is symmetrical to this side. Besides, from Fig.\ref{fig:3} (a), inner peaks rotate faster than outer peaks, and we realize that the galaxy wave is with inhomogeneous angular velocities at different longitudinal distances. We demonstrate this property by analyzing the specific angular variations of inner and outer peak regions with the propagation distance shown in Fig.\ref{fig:3} (b), in which angles of the inner peak regions vary between 0 and 0.919 rad seen from the blue line and angles of the outer peak regions vary between 0 and 0.2857 rad from the red line in the simulation. It can be seen that the angular variation range of the inner peak regions is much larger than that of the outer. Corresponding angular positions in experiment at different propagation distance $z$ are shown as the error bars around simulated lines evaluating the errors to find the average angular positions of peak regions. Those blue and red error bars represent the angular positions of 0.0002$\%$ maximum intensity regions of the inner and the outer layers respectively, at different propagation distance $z$. Fig.\ref{fig:3} (c) shows the corresponding angular rotating velocities of inner and outer peak regions upon propagation respectively as blue and red simulated lines, and experimental results are shown as error bars around simulated lines evaluating the errors to find average angular velocities of peaks. The maximum angular velocity of the inner peak regions is
23.47~rad/m and that of outer peak regions is nearly 3~rad/m. Seen from the large difference of maximum angular velocities, which is nearly an order of 
magnitude, galaxy waves provide great tunable range for inhomogeneity of bi-layer angular velocities.

\begin{figure}
	\centering
	\includegraphics[width=0.55\linewidth]{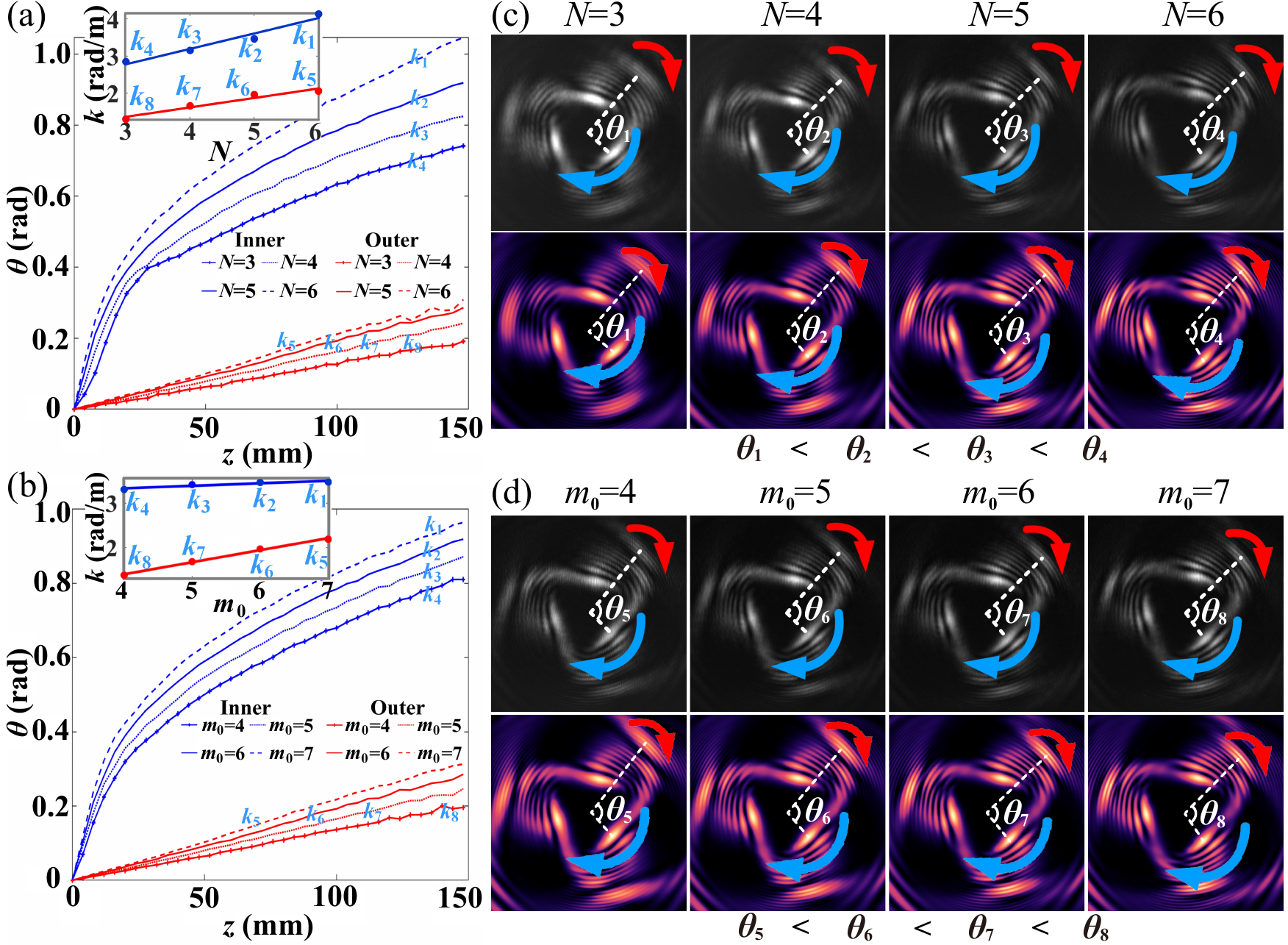}
	\caption{(a,b) Angular position evolutions of inner and outer peak regions of (a) the galaxy waves with fixed $m_{0}$ and $n_{0}$ and $N$ tuned from 3 to 6 and (b) the galaxy waves with $m_{0}$ varying from 4 to 7. The inserts show the evolutions of $N$- and $m_0$-dependent angular velocities as slopes marked correspondingly. (The fitting range for slope is from 40~mm to 148~mm.) (c,d) Experimental and theoretical transverse patterns of the galaxy waves in (a) and (b) at distance of $z=148$~mm. }
	\label{fig:4}
\end{figure}

Besides, galaxy waves have rich tunable parameters for mode control. With different initial index $m_{0}$ or parameter related to the number of superimposed CAVBs $N$, the rotating characteristics of the inner and outer intensity peak regions are tunable. When $m_{0}=6$ and $N$ is tuned from 3 to 6, angular variations of the inner and outer peak regions upon propagation are shown in Fig.\ref{fig:4} (a) as blue lines and red lines in different linestyles respectively. Corresponding slopes $k_{i}$, $i=1,2,\dots,8$, which represent the angular rotating velocities of peak regions within the fitting range 40 to 148~mm, are shown in the illustration. It is noticed that with the increase of $N$, the angular variation ranges expand and angular rotating velocities increase. Moreover, if the parameter related to the number of superimposed CAVBs is fixed to $N=5$ but the initial index $m_{0}$ is tuned from 4 to 7, angular variations of the inner and outer peak regions upon propagation are shown in Fig.\ref{fig:4} (b) as the blue and red lines in different linestyles respectively. Corresponding angular rotating velocities shown as slopes $k_{i}$, $i=1,2,\dots,8$ of the angular variation lines within the fitting range 40 to 148~mm are in the illustration. It's found that with the increase of $m_{0}$, the angular variation ranges also expand and angular rotating velocities increase, but not that much as the case while $N$ increases. Related light fields at $z=148$~mm are shown in Fig.\ref{fig:4} (c) and (d) to indicate the evolution differences caused by $N$ and $m_{0}$ respectively. Therefore, angular velocities of the inner and outer peak regions can be tunable with the designed indices, which increases the flexibility of modes in applications.  

\begin{figure}
	\centering
	\includegraphics[width=0.5\linewidth]{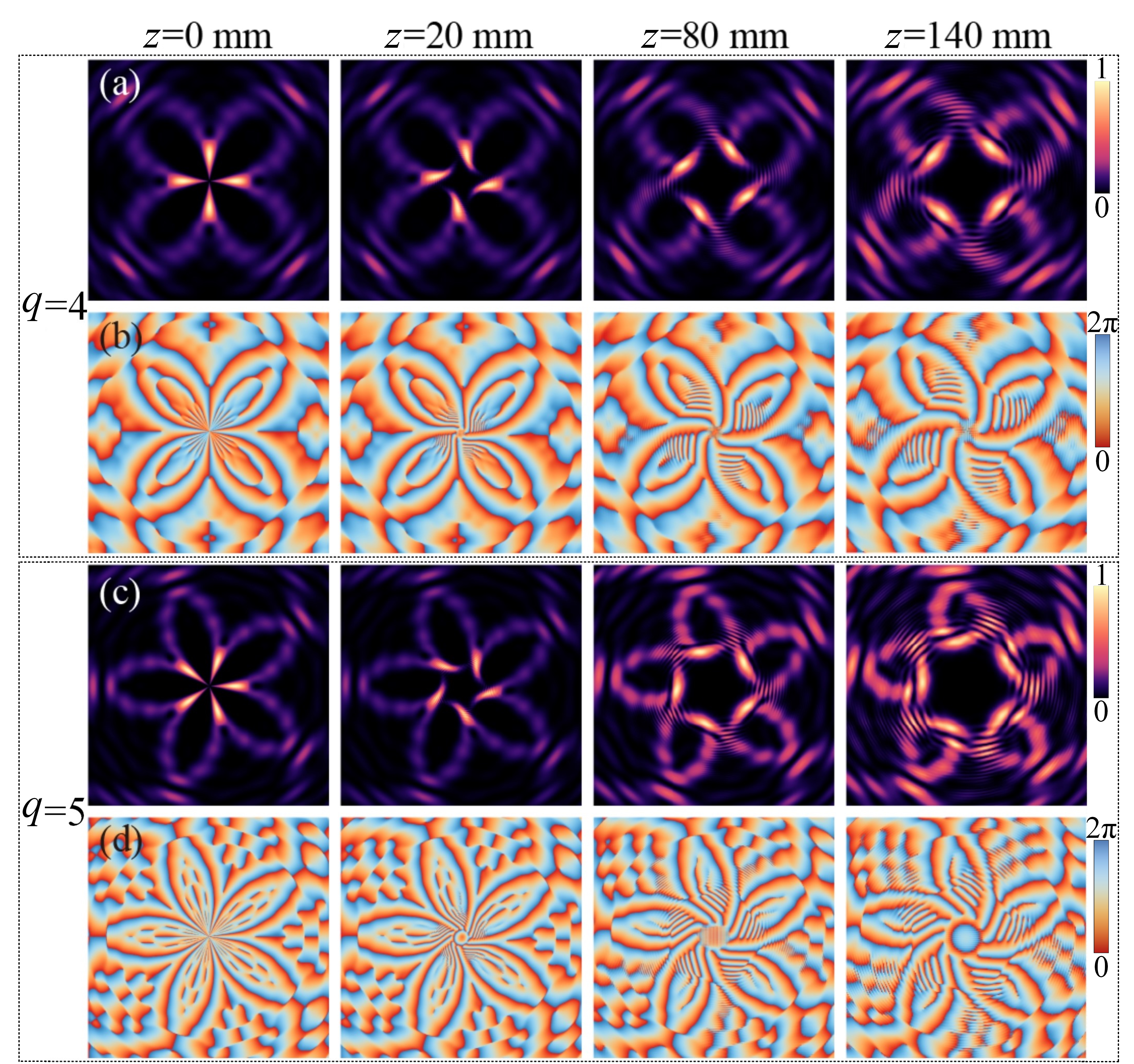}
	\caption{(a,b) Intensity and phase distributions evolved upon propagation of the galaxy wave of $q=4$, and (c-d) the distributions of the galaxy wave of $q=5$, at distances of $z=0$, $20$, $80$, $140$~mm, respectively.}
	\label{fig:5}
\end{figure}

In addition to dynamic characteristics on the angular variation of peak regions, global symmetry of light field distribution can be well controlled by parameter $q$. For example, intensity and phase distributions of the
galaxy wave with $q=4$ at $z=0, 20, 80, 140$~mm are shown in Fig.\ref{fig:5} (a) and (b) respectively, and modes with $q=5$ at $z=0, 20, 80, 140$~mm are also shown in Fig.\ref{fig:5} (c) and (d) respectively. (More detailed evolutions of the intensity distributions with parameter $q=4,5$ are shown in {\color{blue}{Visualization~2,3}} respectively.)

\begin{figure}[H]
	\centering
	\includegraphics[width=0.5\linewidth]{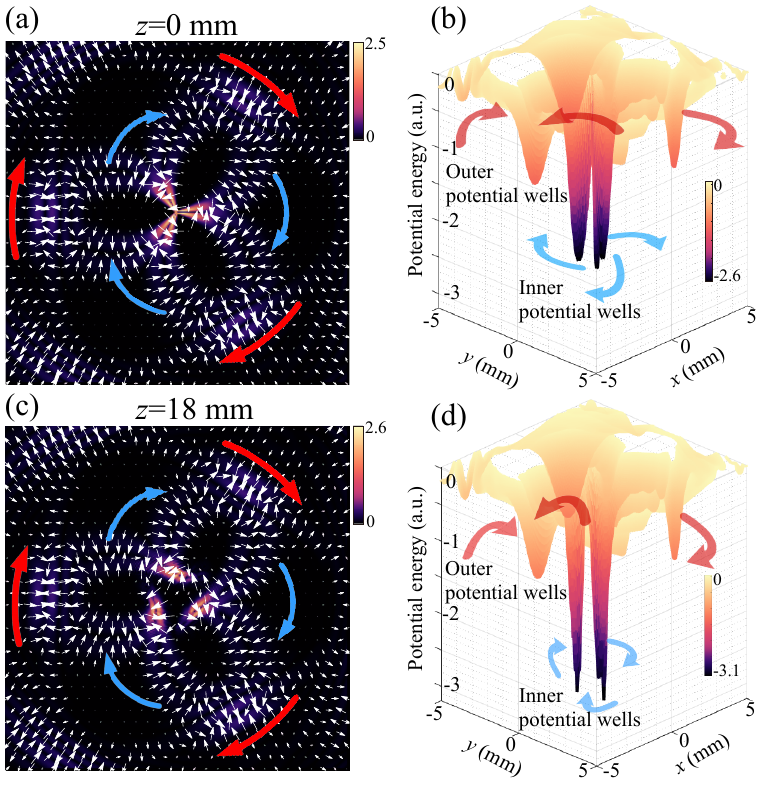}
	\caption{(a) Simulated gradient force at $z=0$ with (b) corresponding potential well, and (c) gradient force at $z=18$~mm with (d) corresponding potential well of the galaxy wave of $q=3$.}
	\label{fig:6}
\end{figure}

Based on the peculiar distribution evolution of peak regions, the intensity gradient of galaxy waves can induce nontrivial gradient force distribution. To demonstrate this claim, we here simulate the gradient force and potential well distribution at different propagation distances with a Rayleigh particle example. Note that for simplicity, we normalize the gradient force and potential well distribution and focus on their relative changes upon propagation in Fig.~\ref{fig:6}. Particles with 0.01~$\mu$m radius and 1.59 refractive index are captured by the galaxy wave ($q$=3) in a liquid with 1.33 refractive index. Particles are trapped at the bottom of potential wells with the combined effect of scattering force $F_{s}$ and gradient force $F_{g}$, respectively shown as~\cite{harada1996radiation}:
\begin{equation}
	\begin{split}
		F_{s}=n_{m}I_{0}\sigma/c,\quad F_{g}=\beta\nabla(\Psi^{2})
		\label{eq:3}
			\end{split}
		\end{equation}
where $n_{m}$ is the refractive index of the liquid, $I_{0}$ is the light intensity, $c$ is the speed of light and $\sigma$ is the scattering cross section. $\sigma=8/3\pi(ka')^{4}a'^{2}(\frac{m^{2}-1}{m^{2}+2})^{2}$, and $\beta=\pi\epsilon_{0}n_{m}^{2}\frac{m^{2}-1}{m^{2}+2}a'^{3}$, where $k=2\pi/\lambda$ ($\lambda$ is the wavelength), $a'$ is the radius of particles, $m$ is the refractive index ratio of the particle and the liquid and $\epsilon_{0}$ is the permittivity of vacuum.

Scattering force and gradient force are proportional to $a'^{6}$ and $a'^{3}$ respectively. Thus gradient force can be designed to be much larger than scattering force by choosing the appropriate radius of particles. Transverse gradient force distribution at $z=0$ is shown in Fig.~\ref{fig:6} (a), based on Eq.~\ref{eq:3}. White arrows point to the inner and outer peak regions, and their lengths represent the relative magnitude of the force. Fig.~\ref{fig:6} (b) shows the corresponding potential well of this galaxy wave in the simulation, whose deep hollows can trap particles. Transverse gradient force distribution at $z=18$~mm is shown in Fig.~\ref{fig:6} (c), and the corresponding potential well is shown in Fig.~\ref{fig:6} (d). The angular variation of the deep hollows reflects the ability of manipulating particles rotating in the transverse dimensions, and the depth variation of the potential well hollows shows the particle manipulative ability along the propagation dimension. Comparing Fig.~\ref{fig:6} (b) and (d), the galaxy wave shows great application potential in particle manipulation in 3D. (The entire evolution is shown in {\color{blue}{Visualization 4}}.)

In conclusion, we design a new self-accelerating structured light named galaxy wave which is angularly symmetric and has radially multilayer intensity peak regions with inhomogeneous rotation angular velocities. Moreover, the peak intensity has a bi-peak distribution upon propagation. The rotating angular velocities of inner and outer intensity peak regions and the mode symmetry can be tuned by adjusting superposition components elaborately. The galaxy wave provides peculiar potential wells for particle manipulation in 3D, due to the transverse inhomogeneous rotating distribution and the longitudinal propagating evolution. For the brilliant flexibility, galaxy waves show huge application potential, especially in optical trapping, manufacturing, and nonlinear optics.

\begin{backmatter}
\bmsection{Funding} 
This work was funded by National Natural Science Foundation of China (61975087).
\bmsection{Disclosures} The authors declare no conflicts of interest.
\bmsection{Data availability} Data underlying the results presented in this paper are not publicly available at this time but may be obtained from the authors upon reasonable request.
\end{backmatter}

% Bibliography

% Full bibliography added automatically for Optics Letters submissions; the following line will simply be ignored if submitting to other journals.
% Note that this extra page will not count against page length
%\bibliographyfullrefs{OSA-journal-template}

%Manual citation list
%\begin{thebibliography}{1}
%\bibitem{Zhang:14}
%Y.~Zhang, S.~Qiao, L.~Sun, Q.~W. Shi, W.~Huang, %L.~Li, and Z.~Yang,
 % \enquote{Photoinduced active terahertz metamaterials with nanostructured
  %vanadium dioxide film deposited by sol-gel method,} Opt. Express \textbf{22},
  %11070--11078 (2014).
%\end{thebibliography}

% Please include bios and photos of all authors for aop articles
\ifthenelse{\equal{\journalref}{aop}}{%
\section*{Author Biographies}
\begingroup
\setlength\intextsep{0pt}
\begin{minipage}[t][6.3cm][t]{1.0\textwidth} % Adjust height [6.3cm] as required for separation of bio photos.
  \begin{wrapfigure}{L}{0.25\textwidth}
    \includegraphics[width=0.25\textwidth]{john_smith.eps}
  \end{wrapfigure}
  \noindent
  {\bfseries John Smith} received his BSc (Mathematics) in 2000 from The University of Maryland. His research interests include lasers and optics.
\end{minipage}
\begin{minipage}{1.0\textwidth}
  \begin{wrapfigure}{L}{0.25\textwidth}
    \includegraphics[width=0.25\textwidth]{alice_smith.eps}
  \end{wrapfigure}
  \noindent
  {\bfseries Alice Smith} also received her BSc (Mathematics) in 2000 from The University of Maryland. Her research interests also include lasers and optics.
\end{minipage}
\endgroup
}{}

\end{document}